\documentclass[twocolumn,pra]{revtex4}
\usepackage{graphicx}
\begin{document}


\title{Experimental Uhrig Dynamical Decoupling Using Trapped Ions}


\author{Michael J. Biercuk}
\email[To whom correspondence should be addressed: ]{biercuk@boulder.nist.gov}
\altaffiliation{\emph{also} Georgia Inst. of Technology, Atlanta, Georgia}
\author{Hermann Uys}
\altaffiliation{\emph{also} Council for Scientific and Industrial Research, Pretoria, South Africa}
\author{Aaron P. VanDevender}
\author{Nobuyasu Shiga}
\altaffiliation{\emph{Present Address} NICT, Tokyo, Japan}
\author{Wayne M. Itano}
\author{John J. Bollinger}
\affiliation{NIST Time and Frequency Division, Boulder, CO, 80305}


\date{\today}

\begin{abstract}
We present a detailed experimental study of the Uhrig Dynamical Decoupling (UDD) sequence in a variety of noise environments. Our qubit system consists of a crystalline array of $^{9}$Be$^{+}$ ions confined in a Penning trap.  We use an electron-spin-flip transition as our qubit manifold and drive qubit rotations using a 124 GHz microwave system.  We study the effect of the UDD sequence in mitigating phase errors and compare against the well known CPMG-style multipulse spin echo as a function of pulse number, rotation axis, noise spectrum, and noise strength.  Our results agree well with theoretical predictions for qubit decoherence in the presence of classical phase noise, accounting for the effect of finite-duration $\pi$ pulses.  Finally, we demonstrate that the Uhrig sequence is more robust against systematic over/underrotation and detuning errors than is multipulse spin echo, despite the precise prescription for pulse-timing in UDD.
\end{abstract}


\maketitle

\section{Introduction}
\indent Quantum information systems will be particularly susceptible to single bit errors --- in stark contrast to their digital classical counterparts --- due in part to the continuous nature of how information is encoded in a qubit.  While this property makes quantum computers similar in some sense to analog computers, the underlying principles of quantum mechanics and the theory of fault-tolerant Quantum Error Correction (QEC) suggest that historical considerations about error accumulation in analog machines are not applicable to future quantum computers \cite{NC2000}.  QEC provides a means to suppress qubit errors to arbitrarily low levels, as required by the needs of a particular algorithm.  In order for QEC to provide a useful benefit, however, single qubit error probability must be sufficiently small that the overhead required to implement the error correction procedure does not increase the likelihood of a logical error more than the capacity of the error correction procedure to reduce the logical error probability.  Suppression of errors below the so-called ``Fault-Tolerant'' threshold ensures a net gain in error reduction by application of QEC protocols.  Estimated values for the threshold vary widely and depend sensitively on the chosen error model and circuit architecture, but are generally in the range $10^{-3}$ to $10^{-6}$ \cite{NC2000}.  Reaching error rates this low is a challenge for any technology, but the requirements of a quantum information processor are further complicated by the desire to suppress errors well below this threshold.  In the standard description, given a qubit error probability $p$, and an error threshold $p_{th}$, the logical qubit error rate becomes $p_{th}(p/p_{th})^{2^{k}}$ for $k$ levels of QEC concatenation (assuming a distance 3 code) \cite{NC2000}.  Therefore we see that the amount by which $p$ is suppressed below $p_{th}$ determines the net benefit in the logical error rate of using a QEC code.
\\
\indent In addition to errors induced by general operational infidelity, qubits are subject to both bit-flip ($X$) and phase-flip ($Z$) logical errors induced by coupling of the qubit to its environment.  These errors arise from two dominant forms of decoherence: longitudinal relaxation, and transverse dephasing, respectively \cite{NC2000}.  For example, a qubit whose Bloch vector has been rotated from a pole to the equator due to longitudinal relaxation has equal probabilities of being measured $\left|\uparrow\right\rangle$ or $\left|\downarrow\right\rangle$ in the qubit basis, commensurate with a 50 $\%$ probability of a bit-flip error.  Qubits may couple to many environmental degrees of freedom, making decoherence --- and especially dephasing --- an insidious source of errors for any quantum computer.  Considerations such as these have motivated a search for methods capable of suppressing qubit decoherence, and hence qubit errors \cite{Zoller2005}.
\\
\indent Dynamical decoupling is a promising and powerful tool in quantum information science for suppressing qubit decoherence \cite{Viola1998,Viola1999, Zanardi1999,Vitali1999, Byrd2003,Khodjasteh2005, Yao2007, Kuopanportti2008, Gordon2008}.  This method involves the repetitive application of control pulses whose aggregate effect is to time-reverse the impact of qubit coupling to environmental noise.   While attractive in its simplicity and predicted effectiveness, the use of dynamical decoupling has to date largely been impractical, as most experimental quantum information systems have suffered from poor operational fidelities in addition to the effects of qubit decoherence.   However, over the last several years trapped ion qubits have demonstrated exceedingly high gate fidelities, leading to possible experimental scenarios in which qubit errors are dominated by environmental coupling rather than poor control.  The incorporation of dynamical decoupling techniques into a quantum information context has been theoretically proposed as a means to suppress qubit errors below the fault-tolerance threshold, but has been, at this point, only sporadically employed for the reasons cited above.
\\
\indent Understanding of how decoupling sequences suppress decoherence is largely based on decades of work in Nuclear Magnetic Resonance (NMR)~\cite{Haeberlen1976, Vandersypen2004}.  It is well known from this community that the standard multipulse spin echo (CPMG, after Carr, Purcell, Meiboom, and Gill) technique is capable of suppressing decoherence associated with slowly fluctuating noise processes.  This is, however, an overly restrictive case in a general quantum informatic context, as environmental coupling may contribute noise which fluctuates rapidly compared with experimentally relevant timescales such as required qubit memory times.  Recent theoretical work has introduced new methods for understanding the influence of arbitrary dynamical decoupling pulse sequences \cite{Uhrig2007,Uhrig2008,Lee2008, Cywinski2008} under realistic experimental conditions, including the use of nonidealized control pulses  \cite{Biercuk2009}, and the presence of noise environments with arbitrary power spectra.  These techniques go far beyond the common Magnus expansions~\cite{Haeberlen1976} employed in NMR-style analyses of the influence of environmental noise on pulse-sequence performance.  However, they require detailed experimental verification in order to evaluate their utility relative to extant analytical techniques.
\\
\indent Based on these new insights, theoretical \cite{Uhrig2007,Uhrig2008,Lee2008, Cywinski2008} and experimental \cite{Biercuk2009} studies have demonstrated that careful manipulation of the pulse-spacing in a dynamical decoupling pulse sequence allows improved suppression of random phase accumulation over standard, periodic pulse sequences \cite{Haeberlen1976, Vandersypen2004, Witzel2007} in high-frequency-dominated noise environments.  In particular, Uhrig developed an analytically optimized pulse sequence, UDD, which was originally designed for qubits subject to an Ohmic noise power spectrum in a spin-boson model \cite {Uhrig2007}.  This model is of interest to a variety of qubit technologies, including electron spins in quantum dots subject to phonon-induced decoherence.  Given $n$ $\pi$-pulses (coherent operations rotating the Bloch vector halfway around the Bloch sphere relative to a defined axis) and a total sequence length $\tau$, Uhrig derived an analytical expression for the relative positions of the $\pi$ pulses such that the sequence strongly suppressed the dephasing effects of noise, with analytical and numerical calculations suggesting the sequence could outperform the well-established CPMG multipulse spin echo~\cite{Haeberlen1976} by orders of magnitude in the high-fidelity (low error) regime.  It was later shown that the UDD sequence could be employed to suppress decoherence in any noise model \cite{Yang2008}, with impressive gains resulting in the context of a spin-bath \cite{Lee2008}.  Together, these results suggest that the use of appropriately engineered dynamical decoupling pulse sequences could provide a route to reaching and exceeding the fault-tolerance threshold for memory operations without the need to restrict our analysis to slowly-fluctuating noise processes.
\\
\indent In order to understand the applicability of dynamical decoupling techniques in future quantum information experiments, we are thus motivated to perform an experimental study of the efficacy of dynamical decoupling pulse sequences in the suppression of decoherence-induced errors.  This study will aim to evaluate theoretical constructs predicting qubit coherence under the influence of dynamical decoupling pulse sequences, and the recent development of novel pulse sequences which promise to provide significant benefits under specific environmental noise conditions.  
\\
\indent In this article we present detailed experimental studies of the performance of dynamical decoupling pulse sequences using an array of trapped-ion qubits as a model quantum memory \cite{Taylor2007, Biercuk2009}.  We drive a $\sim$124 GHz qubit transition in a crystallized nonneutral plasma of $^{9}$Be$^{+}$ ions using a quasi-optical microwave system.  The ions are subjected to environmental noise derived either from intrinsic fluctuations in our magnet system, or artificially synthesized noise in our control system.  We focus on a scenario where dephasing dominates longitudinal relaxation; in this framework, appropriate for most contemporary qubit technologies, dynamical decoupling pulse sequences reduce the accumulation of uncontrolled random phases during qubit free-precession periods.  Qubit decoherence is measured under application of various pulse sequences as a function of sequence length, $\pi$-pulse number $n$, rotation axis, noise spectrum, and noise strength, and is compared against theoretical predictions incorporating experimentally relevant parameters.  We study the performance of the novel UDD sequence and compare against the benchmark multipulse spin echo sequence (of which CPMG is a special case).  Finally, we explore the robustness of these pulse sequences in the presence of common systematic control errors \cite{Zhang2008}.  Our results demonstrate an ability to accurately predict qubit coherence in arbitrary noise environments under the application of arbitrary pulse sequences, and validate theoretical conjectures that careful pulse-sequence engineering can indeed provide significant benefits in error suppression.
\\
\indent We mimic the dephasing processes that may be encountered in solid-state devices by engineering the noise environment in our experiment \cite{Biercuk2009}, but gain high-fidelity control operations, state initialization, and measurement through the use of an atomic system.  As such, our experimental platform behaves as a model quantum system capable of emulating the noise environments of other qubit technologies in the limit of classical dephasing.  This observation suggests that our results are generic across all qubit realizations and that dynamical decoupling techniques should yield benefits in all qubit systems.
\\
\indent This paper is organized as follows.  In Section \ref{sec:thy} we describe the effect of dynamical decoupling pulse sequences on the time-evolution of a qubit subject to random phase noise, introducing the concept of the filter function as originally derived in references \cite{Uhrig2008, Cywinski2008}.  In this section we also introduce the construction of the main sequences under test, and describe their predicted performance in various noise environments.  Section \ref{sec:exptsys} describes our experimental system in detail, including discussion of qubit structure, coherent control techniques, and hardware control.  Experimental results are presented in Section \ref{sec:performance}, comparing the UDD and CPMG pulse sequences in $1/\omega^{4}$ and Ohmic noise power spectra, and demonstrating strong agreement with theoretical fitting functions.  Section \ref{sec:robustness} studies the accumulation of coherent rotation errors due to systematic control infidelities in the application of pulse sequences.  Finally, a discussion of experimental results is presented in Section~\ref{sec:discussion}, and concluding remarks are provided in Section \ref{sec:conclusion}.

\section{\label{sec:thy}Decoherence Under Pulse-Sequence Application}

\indent In order to describe the effects of dynamical decoupling pulse sequences on qubit dephasing, we consider classical phase randomization due to environmental noise.  This is described by a Hamiltonian written as
\begin{equation}
H=\frac{1}{2}[\Omega+\beta(t)]\hat{\sigma}_{Z},
\end{equation}
where $\Omega$ is the unperturbed qubit splitting, $\beta$ is a time-dependent classical random variable \cite{Kuopanportti2008,Cywinski2008}, and $\hat{\sigma}_{Z}$ is a Pauli operator.  Under the influence of this Hamiltonian, a superposition state initially oriented along $\hat{Y}$ evolves in time as $\left|\Psi(t)\right\rangle=\frac{1}{\sqrt{2}}(e^{-i\Omega t/2}e^{-\frac{i}{2}\int_{0}^{t}\beta(t')dt'}\left|\uparrow\right\rangle+e^{i\Omega t/2}e^{\frac{i}{2}\int_{0}^{t}\beta(t')dt'}i\left|\downarrow\right\rangle)$ in the lab frame, with $\left|\uparrow\right\rangle$ and $\left|\downarrow\right\rangle$ the qubit states.  In the frame rotating at $\Omega$, the term $\beta(t)$ randomly rotates the Bloch vector in the equatorial plane, corresponding to the accumulation of a random phase between the qubit basis states.  The presence of any uncontrolled phase increases the probability of a logical $Z$ error, and when the root-mean-squared phase accumulation is $\sim\pi$ the qubit is said to have dephased.  The characteristic timescale for this process is known as $\tau_{\phi}$, and in the absence of relaxation is equivalent to $T_{2}$, generically known as the decoherence time.  This form of decoherence is known as homogeneous dephasing, as it treats the effect of phase randomization on a single qubit or in an ensemble where all qubits experience the same phase evolution.  Alternatively, inhomogeneous broadening occurs when qubits in an ensemble are subject to different phase evolution, or time-ensemble average measurements are made, in either case yielding a characteristic decoherence time known as $T_{2}^{*}\leq T_{2}$ \cite{Haeberlen1976}.

\subsection{The Filter Function Formalism}
\indent Hahn showed that in NMR systems injection of a $\pi$ pulse around $\hat{X}$ (henceforth denoted $\pi_{X}$) halfway through a free-precession period time-reversed the effects of inhomogeneous dephasing \cite{Hahn50}.  Similarly, random phase accumulation can be time-reversed in a homogeneous system, so long as temporal fluctuations in the noise environment are slow compared to the free-precession period.  This technique, known as spin-echo \cite{Hahn50,Haeberlen1976}, forms the basis of modern studies of dynamical decoupling \cite{Vandersypen2004}, in which multiple qubit rotations are used to mitigate random phase accumulation, and hence extend qubit coherence time.
\\
\indent Uhrig \cite{Uhrig2008} and Cywinski et al. \cite{Cywinski2008} studied the coherence of a qubit under the influence of pulse-sequence application in an arbitrary noise environment, and showed that for any $n$-pulse sequence one may write a time-domain filter function, $y_{n}(t)$, with values $\pm1$, alternating between each interpulse free-precession period.  The time-domain filter function accounts for phase accumulation under the influence of external noise, with the accumulation being effectively time-reversed by application of $\pi_{X}$ pulses.  Moving to the frequency domain one may then write $F(\omega\tau)=|\tilde{y}_{n}(\omega\tau)|^{2}$, where $\tilde{y}_{n}(\omega\tau)$ is the Fourier transform of the time-domain filter function, and $\tau$ is the total sequence length.
\\
\indent We modify the time-domain filter function to account for finite-length $\pi$ pulses \cite{Viola2003, Zhang2008, Biercuk2009} by incorporating a delay with length $\tau_{\pi}$ and value zero between free-precession periods.  This approximation assumes that dephasing is negligible during the application of a $\pi$ pulse, consistent with many experimental observations.  Incorporating this delay results in a filter function for an arbitrary $n$-pulse sequence
\begin{eqnarray}
&F(\omega\tau)=|\tilde{y}_n(\omega\tau)|^{2}\nonumber\\
&=|1+(-1)^{n+1}e^{i\omega\tau}+2\sum\limits_{j=1}^n(-1)^je^{i\delta_j\omega\tau}\cos{\left(\omega\tau_{\pi}/2\right)}|^{2} \nonumber\\
&\;
\end{eqnarray}
where $\delta_{j}\tau$ is the time of the center of the $j^{\rm th}$ $\pi_{X}$ pulse, and $\tau$ is the sum of the total free-precession time and $\pi$-pulse times.  The only modification relative to previous work associated with accounting for finite $\tau_{\pi}$ comes in the addition of the cosine term at the end of the expression \cite{Biercuk2009}.  The derivation of this equation appears in the Appendix of this manuscript.
\\
\indent The filter function, $F(\omega\tau)$ provides all information about how an arbitrary pulse sequence will suppress phase accumulation.  Following references \cite{Uhrig2007, Cywinski2008}, for arbitrary noise power spectrum $S_{\beta}(\omega)$, we write the coherence of a state initially oriented along $\hat{Y}$ in a frame rotating with frequency $\Omega$ as $W(\tau)=|\overline{\langle\sigma_{Y}\rangle(\tau)}|=e^{-\chi(\tau)}$, where angled brackets indicate a quantum-mechanical expectation value, the overline indicates an ensemble average, and
\begin{equation}
\chi(\tau)=\frac{2}{\pi}\int\limits_{0}^{\infty}\frac{S_{\beta}(\omega)}{\omega^{2}}F(\omega \tau) d \omega.
\end{equation}
Since the filter function enters the coherence integral as a multiplicative factor of $S_{\beta}(\omega)$, small values of $F(\omega \tau)$ will lead to small values of $\chi(\tau)$, and hence coherence $W(\tau)\sim1$.
\\
\indent This construction describes the influence of an arbitrary dynamical decoupling pulse sequence on qubit coherence in an arbitrary noise environment, and it incorporates experimentally realistic conditions such as the use of noninstantaneous control pulses.  As such it represents a significant departure from the Magnus expansion which has historically been employed in NMR systems to describe spin coherence in a noisy environment.
\\

\subsection{\label{subsec:CPMGvsUDD}The CPMG and UDD Sequences}
\indent The multipulse spin echo sequence was originally developed in the context of NMR systems where inhomogeneous broadening required an ability to refocus the ensemble Bloch vector as it spread out during free precession \cite{Haeberlen1976}.  However, the same sequence is quite effective at suppressing phase randomization due to homogeneous effects (e.g. $\beta(t)$) when noise processes are dominated by low-frequency components (e.g. $S_{\beta}(\omega)\propto1/\omega$).  An $n$-pulse spin echo sequence has all pulses evenly spaced, with the first and last free precession periods half as long as the interpulse free precession periods (Fig.~\ref{fig:pulseseq}c).  In the case where $\left|\Psi_{0}\right\rangle$ (the state vector at the leading edge of the first $\pi$-pulse) is oriented along the direction of the applied microwave field, the sequence is known as CPMG, after its proponents.  For reasons that will be described in a later section, this particular version of the multipulse spin echo sequence is most robust against control error accumulation.
\\
\indent Uhrig showed that manipulation of the relative pulse locations, $\delta_{j}$, for fixed $n$ and $\tau$ leads to modification of $F(\omega\tau)$, providing the ability to tailor the filter function in order to strongly suppress noise with certain spectra.  In particular, he analytically derived an $n$-pulse sequence in which the first $n$ derivatives of $\tilde{y}_{n}(\omega\tau)$ vanish for $\omega\tau=0$.  The resulting sequence, UDD, has $\pi$ pulse locations determined analytically as $\delta_{j}= \sin^{2}[\pi j/(2n+2)]$, for an $n$-pulse sequence (Fig.~\ref{fig:pulseseq}c).
\\
\indent The construction presented above allowed Uhrig to tailor the filter function such that it provided strong suppression of phase accumulation when noise environments possessed significant high-frequency contributions --- a dramatic advance over CPMG.  Uhrig specifically showed \cite{Uhrig2008} that in noise spectra including high-frequency components and a sharp high-frequency cutoff, $\omega_{C}$, such as the Ohmic spectrum ($S_{\beta}(\omega)\propto\omega\Theta(\omega-\omega_{C})$) appropriate for a spin-boson model, the UDD sequence was predicted to yield significant gains in performance relative to CPMG.  By contrast, in the presence of noise with a soft high-frequency cutoff, the error-suppression benefits arising from the form of the filter function for UDD were reduced.  These theoretical results were first validated experimentally in reference \cite{Biercuk2009}.
\\
\indent Numerical simulations showed that by applying the UDD sequence, qubit phase error rates for free-precession time $t$, expressed as $1-W(t)$, could be \emph{orders of magnitude} lower than those achievable using the multipulse CPMG sequence when qubits were subjected to high-frequency-dominated noise.    Accordingly, in this study, we will seek to experimentally characterize the performance of the CPMG and UDD sequences in the presence of low-frequency and high-frequency dominated noise.

\section{\label{sec:exptsys}Experimental System}

    \subsection{Qubit System}
\begin{figure}
  \includegraphics[width=\columnwidth]{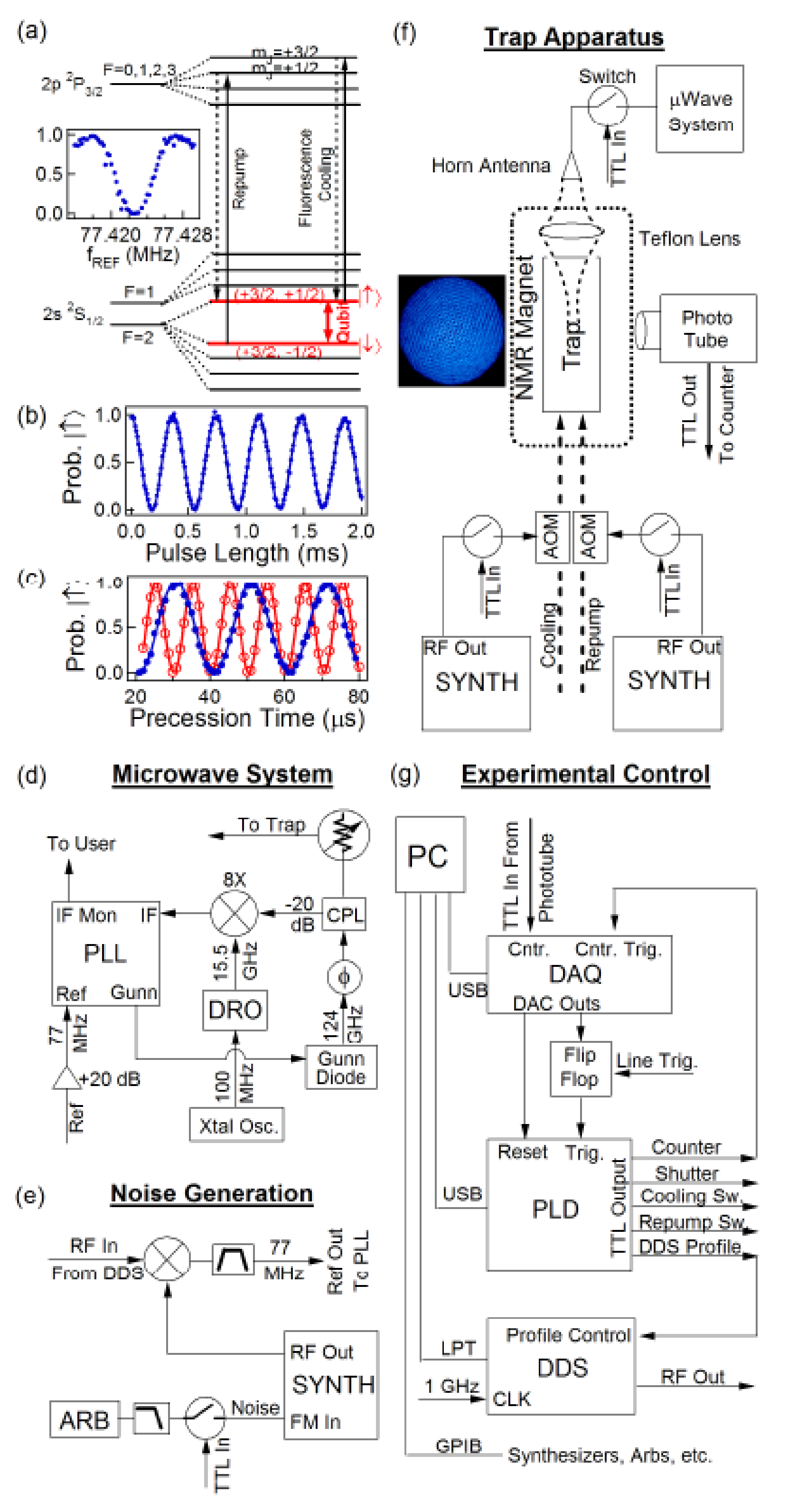}\\
  \caption{\label{fig:schematic} (Color online) Schematic diagram of experimental system.  (a) Qubit level diagram showing relevant transitions and energy levels (see text).  (Inset a) Microwave resonance transition showing Rabi lineshape for square excitation pulses.  Vertical axis is normalized counts.  (b) Rabi oscillations for directly driven qubit transition showing $\tau_{\pi}\sim185\;\mu$s.  (c) Effective Larmor precession measured using a Ramsey interferometer.  Open markers correspond to DDS detuning during free-precession time of 100 kHz, solid markers correspond to 50 kHz detuning. (d) Schematic of microwave generation system.  CPL=coupler.  (e) Schematic of system for generating microwave frequency noise.  (f) Schematic of trap system showing relative positions of hardware elements.  AOM=Acousto-Optic Modulator.  Inset (f) Top-view optical micrograph of an ion array showing hexagonal crystal order, obtained by imaging ion fluorescence parallel to the magnetic field axis. (g)  Schematic of experimental control system.}
\end{figure}

    \indent The qubit employed in our experiments is a ground-state electron-spin-flip transition of $^{9}$Be$^{+}$ ions in the $2s$ $^{2}S_{1/2}$ manifold, ($\left|\uparrow\right\rangle\equiv\left|m_{I} = 3/2, m_{J} = 1/2\right\rangle\leftrightarrow\left|m_{I} = 3/2, m_{J} = -1/2\right\rangle\equiv\left|\downarrow\right\rangle$) \cite{Biercuk2009}.  This transition is highly sensitive to fluctuations in the external magnetic field ($\sim$28 MHz/mT), but has a negligible spontaneous emission rate, making it efficient for studying pure dephasing processes.  We form an ordered qubit array of $\sim$1000 $^{9}$Be$^{+}$ ions in a Penning ion trap \cite{mitchell98, Taylor2007} operated at 4.5 T.  The motion of the ions along the axis of the magnetic field is Doppler cooled to temperatures of order 1 mK \cite{jenm04, jenm05} using $\sim$313-nm UV laser light red-detuned from an atomic transition between the $2s$ $^{2}S_{1/2}$ and $2p$ $^{2}P_{3/2}$ manifolds.  When cooled these ions form 2D or 3D arrays with well defined crystal structure \cite{itaw98, mitchell98, jenm04, jenm05} and ion spacing $\sim$10 $\mu$m.  The ion array rotates rigidly at a frequency $\sim$20 kHz, due to the geometry of the confining electric and magnetic fields.  The rotational frequency is controlled by an external rotating dipole potential \cite{huap98a, huap98b}.  Adjusting the rotation rate of the array allows us to controllably manipulate the dimensionality of the ion crystal (planar vs. 3D).  A schematic of the experimental system is displayed in Fig. \ref{fig:schematic}f.

    \subsection{Initialization, Measurement, and Coherent Control}
    \indent Qubit state initialization into $\left|\uparrow\right\rangle$ occurs via optical pumping by cooling and repump lasers (Fig.~\ref{fig:schematic}a), and state readout is achieved by fluorescence detection on the same cycling transition used for cooling \cite{brel88}.  In our experiments, $\left|\uparrow\right\rangle$ is a bright state that resonantly scatters $\sim$313 nm laser light.  Fluorescence events are detected using a low-dark-count phototube.  In any experimental cycle, after qubit initialization, the cooling/detection laser is pulsed on for a fixed time of 50 ms, fluorescence events are detected, a coherent control pulse sequence is applied, which under conditions of full coherence results in a dark state $\left|\downarrow\right\rangle$, and fluorescence is again detected for 50 ms.  In order to account for optical pumping from $\left|\downarrow\right\rangle\Rightarrow\left|\uparrow\right\rangle$ induced by the detection laser, the second detection cycle is broken into five 10 ms bins, and a linear fit to the number of counts per bin allows extrapolation to a $t=0$ count rate (relative to the start of the detection cycle).  This extrapolated count rate is normalized to pre-control-pulse measurements in order to account for slow fluctuations in laser intensity.  Using these methods we readily achieve preparation and measurement of the dark-state with fidelity greater than 99.5 $\%$, limited by stray-light scatter from the trap electrodes and imperfect $\pi_{X}$ rotations due to magnetic field inhomogeneities.  The best measured fidelity of the dark state prepared by a single $\pi_{X}$ rotation in our system was $99.85$ $\%$ (optimized to maximize microwave coupling to the ions, and minimize inhomogeneities and stray light scatter).
    \\
    \indent Qubit rotations are achieved by directly driving the $\sim$124~GHz electron spin-flip transition using a quasi-optical microwave system.  The use of this technique eliminates the effects of spontaneous emission which are a limiting parameter in Raman-laser-mediated qubit rotations.  Coherent control over this qubit transition was first demonstrated in reference \cite{Biercuk2009}.  We generate 124~GHz microwaves using a Gunn diode oscillator phase-locked to the eighth harmonic of a 15.5~GHz dielectric resonator oscillator (DRO), plus a $\sim$77~MHz IF reference frequency.  The DRO is, in turn, referenced to an ultra-stable 100~MHz quartz oscillator (Fig.~\ref{fig:schematic}d).  The $\sim$77~MHz reference is typically generated using a computer-controlled direct digital synthesizer (DDS) which may be rapidly switched between four preset output profiles (each with fixed phase and frequency) via TTL inputs.
    \\
    \indent Microwaves are launched from a horn antenna and propagate through free space before being focused on the ion array by a Teflon lens (Fig.~\ref{fig:schematic}f).  Pulsed control is afforded via a TTL-controlled p-i-n diode microwave switch that provides $\sim$25 dB of attenuation in the off-state.  The microwave frequency may be detuned while the switch is in the off-state in order to minimize the effects of microwave leakage.  A variable attenuator controls the amount of microwave power output by the horn antenna, and mitigates standing wave reflections from the microwave switch that can disrupt the PLL.  Amplitude stability of our microwave system has been measured to be $>$99.9 $\%$.
    \\
    \indent A fixed-duration microwave pulse is applied to the qubits, and the $\sim$77 MHz reference is swept in order to find the resonant transition frequency, $f_{0}$ (Inset, Fig.~\ref{fig:schematic}a).  This frequency drifts slowly due to flux leakage from our superconducting magnet at a relative rate of $<1\times10^{-8}$ per day.  Additionally, $f_{0}$ can occasionally be seen to shift suddenly throughout the day by up to $\sim$1 kHz, presumably due to the influence of external laboratory equipment in the building.  Careful monitoring of $f_{0}$ allows us to maintain long-term microwave frequency stability relative to the qubit transition of $\sim$1$\times10^{-9}$ over any experimental measurement run.
    \\
    \indent Varying the length of an on-resonance microwave pulse shows Rabi oscillations that decay with a characteristic time of 30 to 40 ms (Fig.~\ref{fig:schematic}b).  Long-time oscillations are best fit with a Gaussian decay envelope arising from microwave amplitude instability, but the complex observed decay pattern suggests the possibility of several competing influences.  Tuning the attenuator upstream from the horn antenna allows us to vary our $\pi_{X}$ time, $\tau_{\pi}$ $\sim70-700$ $\mu$s, while still maintaining reliable $\pi_{X}$ pulses.  Experiments in this study were conducted with $\tau_{\pi}\sim185$ $\mu$s.  Pulse timing resolution of 50 ns corresponds to a systematic pulse timing infidelity of $<4\times10^{-4}$.  A $\pi$ rotation is sensitive to such inaccuracies only to second order.  The infidelity of $\pi$ rotations is $\sim$1$\times10^{-3}$, due to magnetic field and microwave beam inhomogeneities over the ion crystal.
    \\
    \indent Coherent control of qubit phase is demonstrated using a Ramsey interferometer consisting of two $(\pi/2)_{X}$ pulses spaced by a free-precession period.   Microwaves are detuned during the interpulse period, resulting in an effective controlled Larmor precession, as shown in Fig. \ref{fig:schematic}c for detunings of 50 and 100 kHz.  Ramsey interference fringes measured in this way show a characteristic decay time of 1 to 2 ms, depending on the size and configuration of the qubit array.  Ion arrays with multiple planes generally yield shorter decay times due to axial magnetic field inhomogeneities.

    \subsection{Noise Generation}
    \indent We synthesize noise in our experimental system in order to simulate the noise environments of other qubit technologies, and to provide a means to validate the theoretical constructions employed by Uhrig and others \cite{Uhrig2007,Uhrig2008,Lee2008, Cywinski2008}.   Noise synthesis commences with the generation of a desired noise spectrum; all frequency components over a desired range are provided a uniformly distributed random phase $\left[0,2\pi\right]$, and weighted by an envelope function appropriate for the desired noise spectrum (e.g., $1/\omega$).  This spectrum is then Fourier transformed, yielding a numerical time trace whose ensemble-averaged two-time correlation function reproduces the spectrum of interest, and sent to an arbitrary waveform generator (ARB).  The ARB's time-domain voltage output provides a frequency modulation signal for a synthesizer, whose output is mixed with a DDS (Synthesizer: 275 MHz, DDS: $\sim$197 MHz) to yield the $\sim$77 MHz reference to the Gunn diode PLL (Fig. \ref{fig:schematic}e).
    \\
    \indent Our noise synthesis technique involves modulation of the microwave driving frequency rather than injection of magnetic field noise to the magnet bore.  However, modulation of the microwave frequency used for coherent control is equivalent to modulation of the qubit splitting, $\Omega$, relative to a fixed microwave drive, as either technique leads to the same relative phase accumulation between the qubit and the control field.  As such our noise synthesis technique directly mimics the dephasing effects of external fluctuations in the magnetic field (i.e. $\beta(t)$).  Further, this technique faithfully recreates the spectrum of interest in our system, while direct magnetic field-noise injection suffers from, e.g., frequency-dependent shielding in the open-bore NMR magnet.  The noise spectrum is characterized from $10^{-3}$ to $10^{6}$ Hz via direct measurement of a 20 MHz beat-note produced as described above (Synthesizer: 217 MHz, DDS: 197 MHz) using a commercial phase-noise detection system.
    \\
    \indent The noise waveform generated by the ARB and output to the FM port of our synthesizer is blocked by a TTL-controlled RF switch during $\pi_{X}$ pulse application, and is asynchronous with the experimental trigger.  As such, noise is applied only during the interpulse precession period, allowing us to produce high-fidelity $\pi_{X}$-rotations ($\pi_{X}$ fidelity is reduced to $\sim$95 $\%$ if noise is on while driving rotations).  This technique simulates systems that apply very strong, short $\pi_{X}$-pulses, similar to the original prescriptions for dynamical decoupling.
    \\
    \indent In these experiments we set the strength of the injected noise spectrum by adjusting the peak-peak amplitude of the voltage time-trace output by the arbitrary waveform generator, $V_{N}$, where the injected noise power scales as $V_{N}^{2}$.  A representative, smoothed noise spectrum measured for $V_{N}=$0.7 V is shown in the inset to Fig.~\ref{fig:ohmicnoise}a.

    \subsection{Experimental Control}
    \indent Experiments are performed under software control by a PC (Fig.~\ref{fig:schematic}g).  The PC communicates with hardware via various interfaces including USB, parallel port (LPT) and general purpose interface bus (GPIB).  The central hardware component of our setup is a programmable logic device (PLD) that applies TTL control pulses to other hardware, including a DAQ counter for measurement, a microwave switch for pulse application, RF switches controlling repump and cooling lasers, and the profile switch of a DDS.  In addition to programs sent to the PLD, relevant instructions are simultaneously sent to a data acquisition board in order to initialize the counter and other outputs, and GPIB commands are sent to peripheral hardware.
    \\
    \indent For a given experimental sequence, control instructions are sent to the PLD via USB.  A user command synchronized to line frequency triggers the PLD to begin the experiment.  Upon completion of an experiment data are read into software from the data acquisition board, a reset signal is sent to the PLD, new instructions downloaded, and the next experiment commences.  A complete experiment including state initialization, measurement, and the application of the pulse sequence itself takes approximately 200 ms.  A single displayed data point typically consists of the averaged results of 20-50 experiments performed with identical settings.

\subsection{Pulse Sequencing}
\begin{figure}
  \includegraphics[width=\columnwidth]{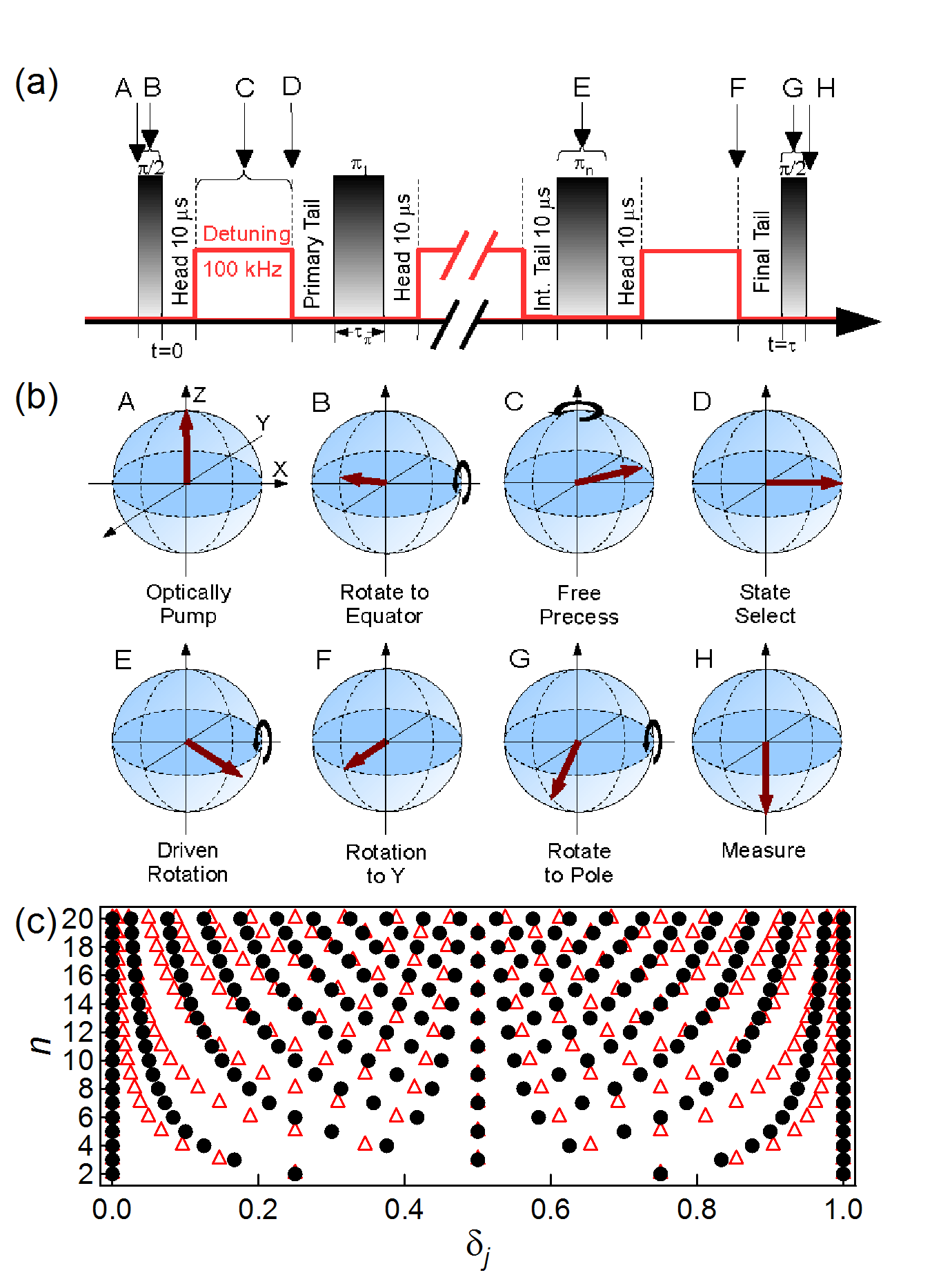}\\
  \caption{\label{fig:pulseseq} (Color online) (a) Schematic pulse sequence showing relevant timescales.  Capitalized letters correspond to specific times and relate to images in lower panel.  Primary and Final Tails set as indicated in the text.  For standard CPMG and UDD Primary Tail: 12.5 $\mu$s, Final Tail: 17.5 $\mu$s.  (b) Bloch sphere representation of key experimental procedures and events in application of a dynamical decoupling pulse sequence.  Curved arrow indicates rotation axis during driving or free precession.  Driven rotations always about $\hat{X}$, free-precession always about $\hat{Z}$.  (c) CPMG (closed markers) and UDD (open markers) fractional pulse locations as a function of pulse number, illustrating increasing divergence between the sequences with larger $n$.}
\end{figure}

\indent A schematic pulse sequence for an arbitrary, $n$-pulse dynamical decoupling sequence is shown in Figs. \ref{fig:pulseseq}a $\&$ b.  The sequence begins with the application of a $(\pi/2)_{X}$-pulse which rotates the Bloch vector to the equatorial plane.  Microwaves are detuned during the interpulse period; application of this detuning results in effective Larmor precession and also serves to reduce the Rabi rate, mitigating unwanted rotations due to microwave leakage.  Short delays called ``Profile Heads (Tails)'' between the trailing (leading) edge of a pulse and the application (removal) of the detuning are employed to precisely control the phase evolution of the qubits.  For instance, with 100 kHz detuning, and the first interpulse precession period set to an integer multiple of 10 $\mu$s, the Primary Tail may be set to 12.5 $\mu$s such that the Bloch vector rotates from the $\hat{Y}$ to the $\hat{X}$ axis.  Varying the length of the Primary Tail therefore allows a form of process tomography in the equatorial plane to be performed, as will be discussed in Section \ref{sec:robustness}.  The Final Tail is set such that in the absence of dephasing the qubits will be rotated to the dark state at the conclusion of the sequence.
\\
\indent When applying CPMG we select the total free-precession time such that each interpulse precession period is an integer multiple of the detuned rotation period, $10\;\mu\rm s = (100\;\rm kHz)^{-1}$.  Accordingly, in the absence of dephasing the Bloch vector makes an integer number of complete rotations about $\hat{Z}$ during each interpulse precession period. Application of UDD generally places no constraints on the length of the interpulse precession period, and at the conclusion of each precession period the Bloch vector ends at a different location on the equator.  However, we find that rounding each precession period to an integer multiple of 10 $\mu$s such that the Bloch vector returns to the same point on the equator before each $\pi_{X}$ pulse makes no difference in our results within measurement uncertainty.
\\
\indent We apply pulse sequences with a few to over 1000 $\pi_{X}$-pulses.  For long sequences we find that rotation errors likely associated with long-time microwave amplitude instability or magnetic field fluctuations build up, yielding minimum measured error rates in dark state probability for 1000 $\pi$ pulses of $\sim$10 to 15 $\%$.  We avoid these errors by focusing primarily on sequences consisting of $n\leq12$ $\pi_{X}$-pulses.  In this regime the minimum error rate for any sequence is $\lesssim1\;\%$.

\section{\label{sec:performance}Sequence Performance in Various Noise Environments}
\begin{figure}
  \includegraphics[width=215pt]{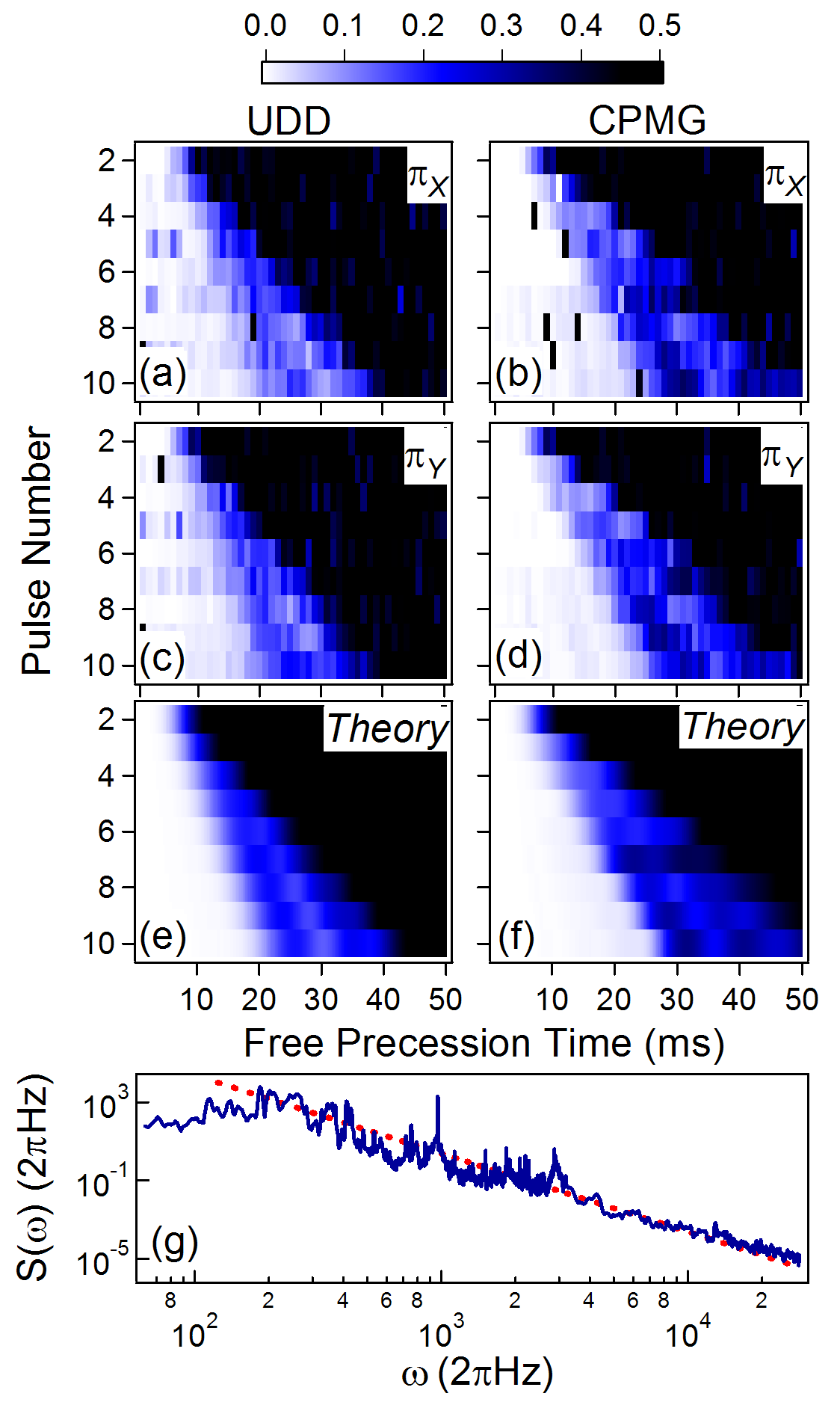}\\
  \caption{\label{fig:pulsenumbers} (Color online) (a-d) Colorscale images of measured qubit decoherence under ambient noise as a function of total free precession time and pulsenumber, $n$, for various sequences.  Colorscale illustrates phase error probability as normalized counts (0.5 = 50 $\%$ probability of ending sequence as $\left|\uparrow\right\rangle$ or $\left|\downarrow\right\rangle$), with light colors corresponding to low qubit errors and black corresponding to full dephasing.  Axis of rotation is designated in the figure.  $\pi_{Y}$ rotations driven by a combination of a $\pi_{X}$ pulse and a net $\pi_{Z}$ rotation during each free precession period.  (e-f) Theoretical fits to qubit decoherence using the filter function as defined in Eq.~(2), and the noise spectrum plotted in panel (g). (g) Measured noise spectrum showing $1/\omega^{2}$ field fluctuations ($S_{\beta}(\omega)\propto1/\omega^{4}$).  Note strong spur at $\omega\sim$ $2\pi\;\times$ 153 Hz.  Dashed line is a guide to the eye representing $S_{\beta}(\omega)\propto1/\omega^{4}$ scaling.}
\end{figure}

\indent We begin our study by applying the CPMG and UDD pulse sequences under the influence of ambient magnetic field noise.  Ambient field noise is measured directly using a field-coil embedded in the bore of our superconducting magnet and a low-frequency spectrum analyzer (Fig.~\ref{fig:pulsenumbers}g).  We find magnetic field fluctuations varying approximately as $1/\omega^{2}$ ($S_{\beta}(\omega)\propto1/\omega^{4}$), with additional sharp spurs, including a prominent feature at $\sim$153 Hz.  While the exact cause of these spurs is undetermined, we believe they are linked to vibrations of the magnet system due to both external perturbations and possibly liquid nitrogen boil-off in the magnet shroud.  We observe the ambient field spectrum to vary slowly and discontinuously in real time, including the overall background level as well as the precise positions and heights of the noise spurs.  The overall scale of the fluctuations in Fig.~\ref{fig:pulsenumbers}g is calibrated by fitting to the decay of Ramsey fringes for a small planar array where effects of inhomogeneous broadening can be neglected.
\\
\indent  We apply the CPMG and UDD sequences under ambient noise while varying the pulse number, showing that in both cases we are able to extend the coherence time with $n$, as predicted by theory (Fig.~\ref{fig:pulsenumbers}).  Qubit decoherence is represented as a colorscale taking values of 0 to 0.5, and corresponds to $\frac{1}{2}(1-W(t))$.  Fully dephased qubits have an equal probability of being measured in either of the two basis states, corresponding to a measured value of 0.5 for error probability.
\\
\indent In our experiments we perform $\pi$ rotations about the $\hat{X}$-axis, or produce an effective rotation about $\hat{Y}$ by combining a $\pi_{X}$ pulse with a net $\pi_{Z}$ rotation during the free-precession period.  The latter technique is similar to the $XZXZ$ decoupling pulse sequences designed to suppress both longitudinal relaxation and transverse dephasing, and may also provide benefits in the context of noise processes that commute with $\sigma_{X}$.  We see that for all times and $n$, these two techniques yield nearly identical results despite the need to round the interpulse precession time to integer multiples of 10~$\mu$s in the case of UDD with $\pi_{Y}$ rotations.  This similarity suggests that at the $10^{-3}$ error level (the limit of our combined $\pi$-rotation and measurement fidelity), the noise suppressing capabilities of UDD are reasonably robust to minor deviations in pulse spacing from analytically derived values.  Numerical simulations confirm that for this noise spectrum, interpulse precession times may be rounded by up to 20~$\mu$s before the minimum error rate consistently exceeds $10^{-3}$.
\\
\indent While the ambient noise spectrum is convenient as it appears naturally in our system, it also provides a useful example of a noise power spectrum dominated by low-frequency components.  In this limit, use of the CPMG sequence is predicted to be an efficient means of phase-error suppression, and the UDD sequence construction should provide no benefits.  As expected, UDD appears to perform similarly to CPMG in the presence of this noise spectrum which has a soft high-$\omega$ cutoff.  Performance in the high-fidelity (low-error) regime is similar between the two sequences, consistent with theoretical expressions for qubit decoherence when accounting for finite $\tau_{\pi}$, and limits imposed by $\pi$-rotation and measurement fidelity.  We note, however, that applying UDD with odd values of $n$ consistently results in a relatively large error rate for times short compared to the coherence time for that sequence ($\sim$10 $\%$ vs. $\sim$1 $\%$).  This appears for both rotation axes and is clearly persistent out to at least $n=10$, but is not predicted by theoretical expressions for qubit decoherence (Figs.~\ref{fig:pulsenumbers}a $\&$ c).  At this point, the observed even-odd asymmetry remains unexplained.
\\
\indent Panels e $\&$ f of Fig.~\ref{fig:pulsenumbers} show fits to theoretical expressions for $\frac{1}{2}\left(1-W(t)\right)$ using two parameters: the overall noise strength ($\alpha$) modifying $S_{\beta}(\omega)$ and entering the integral $\chi(t)$ in Eq.~3, and the relative strength of the 153 Hz spur in the measured noise ($\gamma$).  For these fits we hold $\gamma$ fixed with value 0.15.  Theoretical predictions bear strong resemblance to the experimental data incorporating the measured ambient $S_{\beta}(\omega)$.  As such, fitting factors of order unity are expected.  In the absence of significant $\pi$-pulse imperfections, the same theory applies for $\pi_{X}$ and $\pi_{Y}$ rotations.  Fits are performed for each value of $n$ and show minor variations in extracted $\alpha$ between traces.  The mean value of the fitting factor is $\alpha=$2.3$\pm$0.3, with individual fit uncertainties in the range of 0.1-0.3.  Variations in extracted values of $\alpha$ are consistent with real-time observation of changes in the noise spectrum.  We believe that deviations between our experimental data and fitting functions are dominated by slow as well as discontinuous changes in the ambient noise environment.
\\
\indent At the other extreme of possible noise power spectra, we consider noise dominated by high-frequency contributions for which the simple CPMG sequence is not known to be effective. We mimic the spectrum produced by a spin-boson model in the classical limit for which UDD was originally developed by synthesizing classical noise with an Ohmic spectrum and a sharp-high frequency cutoff \cite{Leggett1987}. Numerical simulations suggest that with this noise spectrum, in the high-fidelity regime UDD is capable of significantly outperforming CPMG, yielding relative gains of several orders of magnitude \cite{Uhrig2007, Cywinski2008}.  
\\
\indent In our experimental system the minimum observable error rates in the high-fidelity regime are limited by measurement infidelity $\sim$0.5~$\%$, making differences in sequence performance indistinguishable in the regime Uhrig originally studied. In order to overcome such limitations we emphasize the differences between these sequences by synthesizing noise with high power relative to the ambient spectrum, thus raising the minimum error rates to above the $\sim$1 $\%$ level.  This approach is well-suited to a study of UDD as follows.   In his original study, Uhrig demonstrated that for fixed resources and an Ohmic noise spectrum, significant error suppression could be gained in the high-fidelity regime at the expense of a slight decrease in coherence time relative to other sequences \cite{Uhrig2007}.  This is manifested as a ``crossover'' point about which the relative performances of UDD and a benchmark sequence (e.g. CPMG) are expected to become inverted.  If this crossover point occurs at low error rates in the high-fidelity regime the benefits of UDD will, in our system, be swamped by measurement infidelity. Uhrig also showed, however, that increasing noise strength can push the crossover point to higher error rates, thus allowing us to enter an experimental regime where differences in sequence performance are easily distinguished.  Thus, in our experimental study we utilize an artificially engineered Ohmic noise spectrum with noise power many orders of magnitude larger than the ambient spectrum.  While our noise does not contain quantum mechanical correlations, the generality of the optimization procedure which led to the discovery of UDD is tested by employing such a noise spectrum.
\begin{figure}
  \includegraphics[width=225pt]{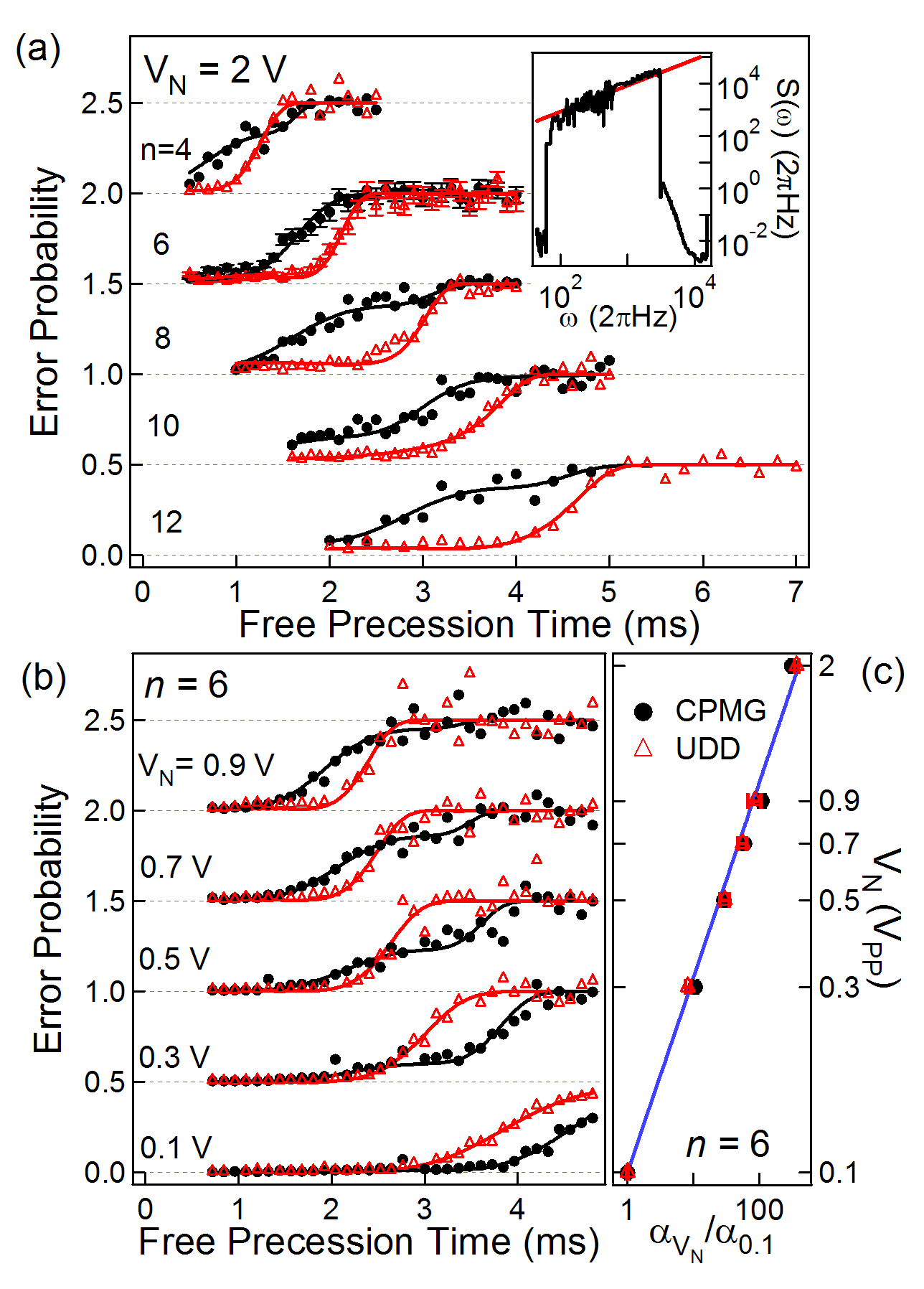}\\
  \caption{\label{fig:ohmicnoise} (Color online) UDD (open markers) and CPMG (solid markers) performance under the application of an Ohmic noise spectrum with a sharp high-$\omega$ cutoff.  (a) UDD and CPMG for fixed noise power and various values of $n$.  Representative error bars presented for $n=6$.  Fits show good agreement with data (see text).  Vertical axis corresponds to an experimental measure of $\frac{1}{2}(1-W(t))$ as in previous figures.  Data offset vertically for clarity.  Inset (a) smoothed, measured ohmic noise spectrum with high-$\omega$ cutoff at $2\pi\;\times$ 500 Hz.  Actual spectrum used for fitting extends from $2\pi\times10$ mHz to $2\pi\times1$ MHz, truncated for visibility.  (b) Evolution of pulse sequences with increasing noise strength, parameterized by $V_{N}$, showing that UDD performance relative to CPMG improves as noise intensity increases.  Data offset vertically for clarity.  (c) Quadratic scaling of extracted fit parameters for various noise strengths as a function of $V_{N}$ on a log-log plot.  All fit parameters normalized to extracted fit parameter for $V_{N}=0.1$ V.  Solid line represents quadratic scaling of the normalized value of $\alpha$ with $V_{N}$.  Data for both CPMG and UDD presented.}
\end{figure}

\indent A smoothed power spectrum with Ohmic scaling, measured for $V_{N}=0.7$ V is displayed in the inset to Fig.~\ref{fig:ohmicnoise}a.  The measured spectral peak at the cutoff of the ohmic spectrum with $V_{N}=0.7$ V is approximately five orders of magnitude larger than the ambient spectrum at the same frequency.  We apply the CPMG and UDD sequences in the presence of this noise power spectrum for $V_{N}=$2 V and display the results for various values of $n$ in Fig.~\ref{fig:ohmicnoise}a.  The UDD sequence clearly and dramatically outperforms CPMG for all measured values of $n$.  Error bars are shown for data with $n=6$, and are derived from the statistical variations in the individual experiments that are averaged to produce a single data point \cite{itaw93}, demonstrating that the traces differ with statistical significance.  As expected, the coherence time may be extended with the addition of $\pi_{X}$ pulses.  Data show good agreement with theoretical predictions for qubit decoherence using the measured noise spectrum and a single fit parameter, $\alpha$, the overall noise scaling.  
\\
\indent We show the performance of UDD and CPMG for various noise strengths, parameterized by $V_{N}$, in Fig.~\ref{fig:ohmicnoise}b.  As predicted in the discussion above, increasing the noise power results in a reversal of the relative performance of the two sequences, with UDD's performance relative to CPMG improving with $V_{N}$.  Again, we find good agreement between data and theory.  In particular, we expect an extracted fit parameter close to unity for $V_{N}=0.7$, the value of ARB output at which the noise power spectrum is measured.  We find  $\alpha_{\mathit{UDD}}=0.88$ and $\alpha_{\mathit{CPMG}}=0.87$.  The small, consistent deviation of these values from one is likely due to the difference in carrier frequency employed in phase noise detection versus noise injection.  
\\
\indent The validity of our fitting procedure is further validated by plotting $\alpha_{V_{N}}/\alpha_{0.1}$ (Fig.~\ref{fig:ohmicnoise}c), the extracted fit parameter for each value of $V_{N}$ normalized to that extracted from $V_{N}=0.1$ V,  as a function of $V_{N}$.  Similarity of the extracted fit parameters for CPMG and UDD is expected as data are taken for both sequences under nominally identical conditions.  These values lie along a straight line with slope two on a log-log plot, consistent with noise power that scales quadratically with $V_{N}$ (noise power scales as the square of the field fluctuations).  As an example, for $V_{N}$ = 0.5 V, we expect $\alpha_{0.5}/\alpha_{0.1}=(0.5\;/\;0.1)^{2}=25$, and obtain for UDD $30\pm5$.  The strong agreement between data and theory is indicative of the general capabilities of the presented theoretical framework.

\section{\label{sec:robustness}Sequence Robustness to Systematic Errors}
\indent Systematic errors associated with quantum control mechanisms promise to be a significant challenge for realizing a quantum information system \cite{NC2000}.  It is therefore imperative to study the robustness of any dynamical decoupling pulse sequence against errors associated with the application of imprecise $\pi_{X}$-pulses \cite{Khodjasteh2005,Zhang2008}.  Further, as quantum information systems will require that dynamical decoupling techniques be applied to arbitrary initial states, it is vital that the sequence performance be known for a variety of input states, information gained through a procedure known as process tomography.  In this section we present measurements of the effect of rotation errors on UDD and CPMG sequence performance, characterized for a variety of input states on the equatorial plane of the Bloch sphere (following the initial $(\pi/2)_{X}$ pulse).

\begin{figure}
  \includegraphics[width=250pt]{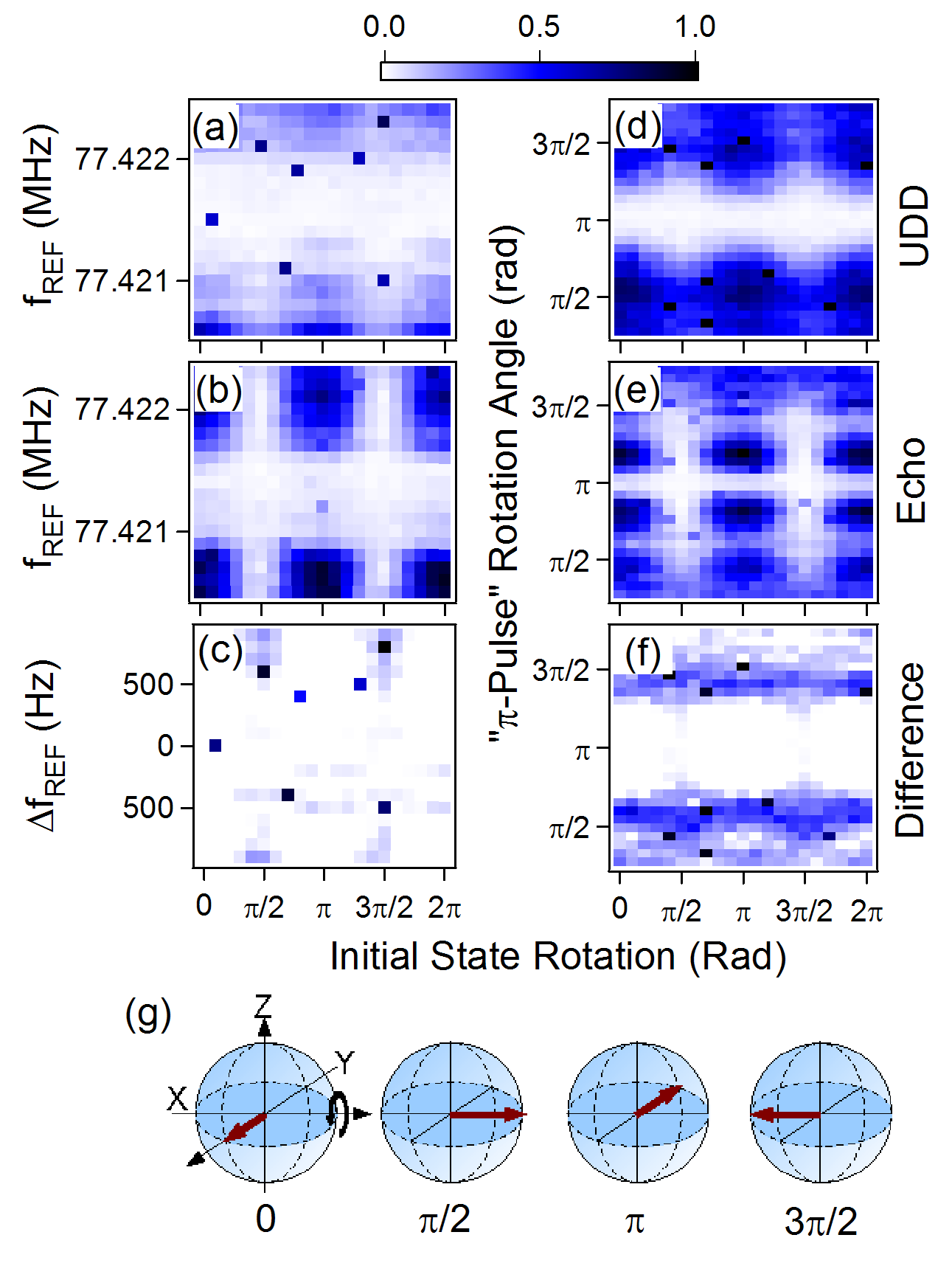}\\
  \caption{\label{fig:robustness} (Color online) Robustness of $n=6$ spin echo and UDD pulse sequences to systematic rotation errors for a variety of input states, measured for free-precession time 8 ms and ambient noise. (a-b) Colorscale representation of qubit rotation to $\left|\uparrow\right\rangle$ as a function of microwave reference frequency (centered on $f_{0}$) and initial state.  Dark colors indicate rotation errors.  (c) Difference between UDD and echo results for above parameters.  White indicates regions where UDD outperforms spin echo.  (d-e) Colorscale representation of normalized counts as a function of actual ``$\pi$-pulse'' rotation angle ($\pi=$185 $\mu$s) where $\tau_{\pi}$ is manipulated to mimic systematic rotation errors.  (f)  Difference between UDD and echo results for ``$\pi$-pulse'' rotations, white shows regions where UDD outperforms echo.  Randomly located black dots correspond to laser dropouts during data collection. (g) Bloch sphere representations of several relevant starting states used in colorscale images above.}
\end{figure}

\indent In NMR spectroscopy, the CPMG sequence is known to be particularly robust to over- and under-rotation errors associated with inaccurate $\pi_{X}$ pulses \cite{Haeberlen1976}.  The sequence, however, is highly sensitive to the initial state on the equatorial plane, providing robust performance only when the Bloch vector is aligned with the direction of the applied microwave field. In this situation, with $\left|\Psi_{0}\right\rangle=\left|X\right\rangle$, the application of a $\pi_{X}$ pulse yields only a small excursion of the Bloch vector about $\hat{X}$.  By contrast, initializing $\left|\Psi_{0}\right\rangle=\left|Y\right\rangle$, the application of $\pi_{X}$ pulses produces large excursions of the Bloch vector along the prime meridian of the Bloch sphere, and provides little resistance to rotation errors.  Similarly, due to the nature of the pulse spacing in UDD, when the interpulse precession periods are unconstrained (and unrounded) we find that for each $\pi_{X}$-pulse the Bloch vector assumes a different location on the equator and rotates in a plane parallel to that defined by the prime meridian.  As such the UDD sequence does not immediately appear as if it will provide the same robustness to pulse errors as does CPMG.
\\
\indent We perform process tomography for $n=6$ $\pi_{X}$ pulse UDD and multipulse spin echo under the influence of ambient noise. The initial state is varied by setting the ``primary tail'' (see Fig. \ref{fig:pulseseq}a) in our pulse sequence such that the Bloch vector rotates by a fixed amount prior to the application of the first $\pi_{X}$ pulse.  A rotation about $\hat{Z}$ by angle $\pi/2$ corresponds to the Bloch vector initially pointing along $\hat{X}$, parallel to the direction of the microwave field vector.  In these experiments we induce rotation errors by two distinct means: changing the length of the $\pi_{X}$ pulse relative to the actual $\tau_{\pi}$, and detuning the microwave frequency from resonance.  These two error mechanisms inject systematic over- and under-rotations as well as phase rotations.
\\
\indent Our measurements clearly demonstrate that the UDD sequence provides significantly more robustness to errors over a wider variety of states than does multipulse spin-echo (Fig.~\ref{fig:robustness}).  We observe that the CPMG sequence (initial state rotated $(\pi/2)_{Z}$) provides the best performance under both forms of systematic error, yielding error rates less than about 10 $\%$ for deviations up to 50 $\%$ of $\tau_{\pi}$ or approximately half a resonance linewidth. The multipulse spin echo sequence, however, is far less robust to error for other initial states on the equator. By contrast, the UDD sequence shows strong robustness across a variety of initial states. For example, for a $\sim$14 $\%$ deviation in $\tau_{\pi}$ and an initial rotation $\pi_{Z}$, the UDD sequence shows a final state rotation error of $\sim$10 $\%$ while the multipulse spin echo has been overrotated by $\sim$70 $\%$. Similarly, a systematic detuning of 500 Hz (1/4 linewidth) yields an accumulated error of $\sim$8 $\%$ for UDD while multipulse echo experiences an error of $\sim$50 $\%$. Panels c $\&$ f of Fig.~\ref{fig:robustness} show the difference between results for the two sequences, with white indicating conditions under which UDD outperforms multipulse spin echo.
\\
\indent These results are striking, as they indicate that despite the precise prescription for the conditions under which UDD is shown to work well, it appears to be quite robust to systematic errors, and even more robust than standard spin echo for a variety of input states. Errors may be suppressed to the few-percent level with UDD, even for sizeable systematic rotation errors. While such errors are not due to a loss of phase coherence, they represent large sources of logical phase-flip and bit-flip errors in the case where dynamical decoupling pulse sequences are incorporated into more complex quantum algorithms.

\section{Discussion}\label{sec:discussion}
\indent A key requirement for the adoption of any quantum control technique is the ability to accurately predict a qubit's state after the application of that technique.  In the context of dynamical decoupling, that requirement is translated to an ability to accurately predict qubit coherence under pulse sequence application.  Historically, this capability was derived from experimentally validated procedures borrowed from NMR, but significant limitations prevented general applicability.
\\
\indent The sensitivity of quantum information processing tasks to dephasing-induced errors requires more accurate techniques for predicting qubit coherence, and more general error-suppression strategies than those introduced in NMR.  In this work we have described a general formalism to predict qubit coherence under the application of dynamical decoupling pulse sequences with noninstantaneous control pulses.  This  formalism applies broadly to arbitrarily spaced pulses, and accounts for the influence of arbitrary noise spectra.  Moreover, we have experimentally implemented UDD, a novel dynamical decoupling pulse sequence constructed using the above formalisms in order to provide enhanced error-suppression performance in the presence of fast noise.
\\
\indent Our experimental results validate the conditions under which the UDD sequence is predicted to outperform CPMG, and in so doing, validate the underlying theoretical constructs predicting qubit coherence.  We have studied the relative performance of these sequences under the approximate limiting cases of low-frequency-dominated ($S_{\beta}(\omega)\propto 1/\omega^{4}$) and high-frequency-dominated ($S_{\beta}(\omega)\propto\omega\Theta(\omega-\omega_{C})$) noise power spectra, in both cases obtaining good agreement between measurements and theoretical calculations.  We expect sequence performance to be intermediate for noise power spectra falling between these approximate limiting cases, as is likely for experimental systems which suffer from a variety of noise processes.
\\
\indent Our measurements demonstrate that in the presence of noise power spectra with significant high-frequency contributions, the UDD sequence is capable of outperforming the benchmark multipulse spin echo (CPMG).  These measurements employ artificially synthesized noise with very high power in order to overcome measurement limitations, but there is no reason to believe that the agreement between theory and experiment would break down in the low-error regime in the presence of weaker noise, or in a system with improved measurement fidelity.  Accordingly, we believe that the UDD pulse sequence --- which requires no more control resources than CPMG --- may provide benefits of many orders of magnitude in the presence of appropriate noise spectra.
\\
\indent We remind the reader that measurements are limited in the high-fidelity regime by readout infidelity.  We emphasize that the specific observed error rates or coherence times are not the main results of our study.  Instead, the precise ability of theory to replicate coherence lineshapes for various pulse sequences, pulse numbers, noise spectra, and noise strengths demonstrates the general applicability of these techniques.
\\
\indent Even in the case of low-frequency-dominated noise, our measurements and previously presented numerical simulations~\cite{Uhrig2007,Uhrig2008,Lee2008, Cywinski2008} suggest that the performance of the UDD sequence is similar to that of CPMG (rather than substantially worse), except in the measured coherence \emph{time}.  For long free-precession times, in low-frequency-dominated noise environments CPMG indeed outperforms UDD.  In a computational setting, however, the accumulation of error near the $1/e$ decay level is already many orders of magnitude larger than that tolerable in a fault-tolerant context.  Accordingly, from a practical perspective, UDD may remain an attractive choice for an experimentalist with an ill-defined or discontinuously variable noise environment, even if the noise is typically dominated by low-frequency components.  This suggestion is supported by the measured robustness of the UDD sequence to coherent rotation errors and timing inaccuracies across a variety of relevant input states.  Further, this robustness may make UDD an attractive sequence for NMR spectroscopists.
\\

\section{Conclusion}\label{sec:conclusion}
\indent In this article we have provided a detailed experimental study of the application of dynamical decoupling pulse sequences in a model atomic system. In particular, we have compared the performance of an analytically optimized sequence, UDD, to standard multipulse spin echo in a variety of noise environments. Our measurements have verified theoretical predictions that under certain experimental conditions, UDD can dramatically outperform the familiar CPMG sequence. Further, we have been able to accurately predict qubit decoherence using an extension of theoretical constructs that incorporates nonzero $\tau_{\pi}$. Fits using our modified theoretical expressions show strong agreement with experimental data as a function of free-precession time, $\pi$ pulse number, noise spectrum, and noise strength.
\\
\indent In addition to verifying a large body of literature on the effects of dynamical decoupling generally, and the performance of the UDD sequence specifically, our experiments solidify the utility of dynamical decoupling as a viable means of error suppression for quantum information processing. The high degree of agreement observed between data and theory, coupled with previous theoretical calculations, suggest that the use of optimized dynamical decoupling pulse sequences has the ability to suppress qubit phase-flip errors to near or below the fault-tolerance threshold without an exponential increase in resources. 
\\
\indent Future experiments will explore the benefits of concatenation of optimized pulse sequences in suppressing phase errors, and the use of composite pulses to improve control fidelity. In addition, we will study more general noise models including the effects of longitudinal relaxation and quantum mechanical correlations by the coherent excitation of phonon modes in our ion crystal.
\\
\appendix*
\section{Derivation of filter function accounting for finite-length $\pi$ pulses}\label{sec:appendix}

\indent In this appendix we present the derivation of a modified filter function that accounts for the influence of finite $\tau_{\pi}$, as used in Eq.~(1).
\\
\indent We introduce the following convention: Assume the total dynamical decoupling sequence length, delays plus the sum of all $\pi$-pulse durations, is $\tau$.  If the center of the $j$'th $\pi$-pulse occurs at time $t=t_j$ then let $\delta_j=t_j/\tau$, $\delta_{n+1}=1$ and let each $\pi$-pulse have duration  $\tau_\pi=\delta_\pi\tau$.
The filter function is given by the modulus squared of  $y_n(\omega\tau)$ which is defined as
\begin{widetext}
\begin{equation}
y_n(\omega\tau)=i\omega\int_{-\infty}^{\infty} s(t^\prime)e^{i\omega t^\prime}dt^\prime
\end{equation}
\begin{equation}
s(t)=\left\{\begin{array}{ll}
0, &  \;\;\textnormal{for}\;\;t\leq0 \\
(-1)^j, & \;\;\textnormal{for}\;\; (\delta_j+\delta_\pi/2)\tau\leq t\leq(\delta_{j+1}-\delta_\pi/2)\tau\;\;\;\;\;\;
\textnormal{(between $\pi$-pulses)}\\
0, & \;\;\textnormal{for}\;\; (\delta_j-\delta_\pi/2)\tau< t<(\delta_{j}+\delta_\pi/2)\tau\;\;\;\;\;\;\;\;\;\;\textnormal{(during $\pi$-pulses)}\\
0, &  \;\;\textnormal{for}\;\; t\geq\delta_{n+1}\tau
\end{array}\right.
\end{equation}
then
\begin{eqnarray}
y_n(\omega\tau)&=&\left\{ -\left[e^{i\omega(\delta_1-\delta_\pi/2)\tau}-1\right] + \left[e^{i\omega(\delta_2-\delta_\pi/2)\tau}-e^{i\omega(\delta_1+\delta_\pi/2)\tau}\right]\right.\nonumber\\
&\;&\hspace{1cm}\left.-\left[e^{i\omega(\delta_3-\delta_\pi/2)\tau}-e^{i\omega(\delta_2+\delta_\pi/2)\tau}\right]+...\;\;\;
+(-1)^n\left[e^{i\omega(\delta_{n+1}-\delta_\pi/2)\tau}-e^{i\omega(\delta_n+\delta_\pi/2)\tau}\right]  \right\}\nonumber\\
&=&
\left\{1 -e^{i\omega\delta_1\tau}\left[e^{i\omega\delta_\pi/2\tau}+e^{-i\omega\delta_\pi/2\tau}\right] + e^{i\omega\delta_2\tau}\left[e^{i\omega\delta_\pi/2\tau}+e^{-i\omega\delta_\pi/2\tau}\right]\right.\nonumber\\
&\;&\hspace{1cm}\left.-e^{i\omega\delta_3\tau}\left[e^{i\omega\delta_\pi/2\tau}+e^{-i\omega\delta_\pi/2\tau}\right]+...\;\;\;
+(-1)^{n+1}ne^{i\omega\delta_{n+1}\tau} \right\}\nonumber\\
&=&
1+(-1)^{(n+1)}e^{i\omega\delta_{n+1}\tau}+2\sum\limits_{j=1}^n(-1)^je^{i\omega\delta_j\tau}\cos{(\omega\tau_\pi/2)}
\end{eqnarray}
\end{widetext}

\begin{acknowledgments}
The authors thank L. Cywinski, S. Das Sarma, V. V. Dobrovitski, X. Hu, E. Knill, S. Lyon, G. Uhrig, and W. Witzel for useful discussions.  We also thank Y. Colombe, D. Hanneke, and D. J. Wineland for their comments on the manuscript.  We acknowledge research funding from IARPA and the NIST Quantum Information Program.  M. J. B. acknowledges fellowship support from IARPA and Georgia Tech, and H. U. acknowledges support from CSIR.  This manuscript is a contribution of the US National Institute of Standards and Technology and is not subject to US copyright.
\end{acknowledgments}

\bibliography{scibib}
\end{document}